\documentclass[aps,pre,twocolumn,groupedaddress]{revtex4-1}

\usepackage{xcolor}

\usepackage{amsfonts,amssymb,amsmath} 
\usepackage{color} 
\usepackage{graphicx} 
\usepackage{epstopdf,mathrsfs} 
\usepackage[colorlinks,linktocpage,linkcolor=blue]{hyperref}

\begin{document}

\title{Langevin picture of L\'{e}vy walk in a constant force field}


\author{Yao Chen}
\author{Xudong Wang}
\author{Weihua Deng}

\affiliation{School of Mathematics and Statistics, Gansu Key Laboratory
of Applied Mathematics and Complex Systems, Lanzhou University, Lanzhou 730000,
P.R. China}

\begin{abstract}

L\'{e}vy walk is a practical model and has wide applications in various fields. Here we focus on the effect of an external constant force on the L\'{e}vy walk with the exponent of the power-law distributed flight time $\alpha\in(0,2)$. We add the term $F\eta(s)$ ($\eta(s)$ is the L\'{e}vy noise) on a subordinated Langevin system to characterize such a constant force, being effective on the velocity process for all physical time after the subordination.
We clearly show the effect of the constant force $F$ on this Langevin system and find this system is like the continuous limit of the collision model. The first moments of velocity processes for these two models are consistent. In particular, based on the velocity correlation function derived from our subordinated Langevin equation, we investigate more interesting statistical quantities, such as the ensemble- and time-averaged mean squared displacements.
Under the influence of constant force, the diffusion of particles becomes faster. Finally, the super-ballistic diffusion and the non-ergodic behavior are verified by the simulations with different $\alpha$.

\end{abstract}


\maketitle

\section{Introduction}
The movements of particles are often in an external potential.
The particles in complex disordered systems influenced by the external force generally  exhibits anomalous diffusion behavior \cite{JeonMetzler:2012,EuleFriedrich:2009,CairoliBaule:2015,MagdziarzWeronKlafter:2008,ChenWangDeng:2017,FedotovKorabel:2015}, which is characterized by the nonlinear evolution in time of the ensemble-averaged mean squared displacement (EAMSD), i.e.,
\begin{equation}\label{1}
\langle(\Delta x(t))^2\rangle=\langle[x(t)-\langle x(t)\rangle]^2\rangle \,{\propto}\,t^\beta  \quad  (\beta\neq1),
\end{equation}
which represents subdiffusion for $0<\beta<1$ and superdiffusion for $\beta>1$; for the case $\beta=2$, it is called ballistic diffusion.

Just as the brackets $\langle\cdots\rangle$ in Eq. \eqref{1} shows, the EAMSD is the statistical average over a large amount of stochastic realizations and thus it is not easy to be measured in experiments. In 1914, Nordlund determined the diffusion coefficients of the traced droplets from separate analysis of single trajectory \cite{Nordlund:1914}. After that, people started to pay attention to evaluate the recorded time series in terms of the time-averaged mean squared displacement (TAMSD) \cite{MetzlerJeonCherstvyBarkai:2014}. Especially, for the case that an external force makes a nonzero mean value of the displacement, the TAMSD is defined as
\cite{BurovJeonMetzlerBarkai:2011,AkimotoCherstvyMetzler:2018}:
\begin{equation}\label{TAdefination}
\begin{split}
  \overline{\delta^2(\Delta)}
  &=\frac{1}{T-\Delta}\int_0^{T-\Delta}  [x(t+\Delta)-x(t)  \\
  &~~~    -\langle x(t+\Delta)-x(t)\rangle]^2dt,
\end{split}
\end{equation}
where $\Delta$ is the lag time; it separates the displacement between trajectory points and is much shorter than measurement time $T$ to obtain good statistical properties. The single particle tracking techniques have been widely employed to study diffusion of particles in living cell \cite{GoldingCox:2006,WeberSpakowitzTheriot:2010,BronsteinIsraelKeptenMaiTalBarkaiGarini:2009}.

One representative model to describe diffusion phenomenon is L\'{e}vy walk \cite{ZaburdaevDenisovKlafter:2015,SanchoLacastaLindenbergSokolovRomero:2004,RebenshtokDenisovHanggiBarkai:2014,ZaburdaevDenisovHanggi:2013}. This model characterizes particle motion with finite velocity, and seems to be more reasonable and originally characterized by coupled continuous time random walk (CTRW) \cite{ShlesingerKlafterWong:1982,KlafterBlumenShlesinger:1987,Zaburdaev:2006,ZaburdaevDenisovKlafter:2015}, in which the probability density functions (PDFs) of jump length and flight time are coupled through a constant velocity. Depending on the power exponent $\alpha$ of PDF of flight time, L\'{e}vy walk could describe ballistic diffusion ($0<\alpha<1$), sub-ballistic superdiffusion ($1<\alpha<2$), and normal diffusion ($\alpha>2$). Later, the equivalent Langevin picture of L\'{e}vy walk is presented in Refs. \cite{EuleZaburdaevFriedrichGeisel:2012,WangChenDeng:2019}. Since the L\'{e}vy walk model is especially practical and can depict multiple types of diffusion behaviors, it has wide applications in various fields, not only in the tracking studies of animals or humans \cite{Nathan.etal:2008}, but also the anomalous superdiffusion of cold atoms in optical lattices \cite{KesslerBarkai:2012}, endosomal active transport within living cells \cite{ChenWangGranick:2015}, etc.

Based on the statistical quantities---EAMSD and TAMSD, the ergodic behavior, an important property of stochastic process, can be investigated. The free L\'{e}vy walk has shown a very special phenomenon about the ergodic property. It says that the free L\'{e}vy walk presents ``ultraweak'' non-ergodic behavior for the power exponent $\alpha\in(1,2)$ \cite{GodecMetzler:2013}. The word ``ultraweak'' means that the TAMSD and EAMSD only differ by a constant factor independent of the lag time $\Delta$ for a long measurement time. While for the case $\alpha\in(0,1)$ \cite{FroembergBarkai:2013} with divergent first moment of flight times, the TAMSD is not self-averaged when the measurement time tends to infinity, although the ensemble-averaged TAMSD only differs with EAMSD by a constant factor.

In this paper, we pay attention to how the L\'{e}vy walk model reacts to an external constant force in physical time. There has been a collision model proposed in Ref. \cite{BarkaiFleurov:1998} describing such a motion. It assumes that the velocity of a free test particle is changed after the collision with the surrounding bath particles.
If the model is under the influence of a constant force $F$, it takes effect between two successive collisions; the first moments of velocity and displacement of the model are derived in Ref. \cite{BarkaiFleurov:1998}.
Here, we deal with this problem by building its Langevin picture, and detailedly discuss the way of adding the external force which can make an effect on the process for the whole physical time. We establish a Langevin system coupled with a subordinator to characterize this process and then find it looks like a continuous limit of the collision model.
The first moments of velocity and displacement are verified to be consistent in these two models. Besides, we can make use of the advantage of Langevin equation to calculate the velocity correlation function of the concerned process, which has not been derived before.

The method to obtain the velocity correlation function is technical, where the four-point PDF of the inverse subordinator has to be used.
We establish the relationship between the subordination \cite{BauleFriedrich:2005} and the renewal theory \cite{GodrecheLuck:2001}, which enhances the understanding of these theories and greatly simplifies the calculations.
Finally, the velocity correlation function is found to be a sum of two independent terms. Then the EAMSD and TAMSD can be directly obtained by applying the generalized Green-Kube formula \cite{DechantLutzKesslerBarkai:2014,MeyerBarkaiKantz:2017}.


The structure of this paper is as follows. In Sec. \ref{two}, we review the subordinator and the inverse subordinator, and then build their relationship with the renewal theory. In Sec. \ref{three}, we establish the Langevin picture of L\'{e}vy walk in the presence of a constant force in physical times. Then we calculate the first moment, the velocity correlation function, EAMSD and TAMSD in Sec. \ref{four}--\ref{six}, respectively.
Finally, we make the summaries in Sec. \ref{seven}. The detailed derivations of some results are presented in Appendix.

\section{Subordinator and inverse subordinator}\label{two}
Subordinator is a non-decreasing L\'{e}vy process \cite{Applebaum:2009} and can be regarded as a stochastic model of time evolution. Especially, it helps to describe different kinds of subdiffusion \cite{Fogedby:1994,MetzlerKlafter:2000-2,ChenWangDeng:2018-2} or superdiffusion \cite{FriedrichJenkoBauleEule:2006-2,EuleZaburdaevFriedrichGeisel:2012,WangChenDeng:2019} processes when coupled with an overdamped or underdamped Langevin equation. In order to characterize the power-law distributed flight time of L\'{e}vy walk, the subordinator $t(s)$ in this paper is taken to be $\alpha$-dependent ($0<\alpha<2$) one with the characteristic function $g(\lambda,s):=\langle\textrm{e}^{-\lambda t(s)}\rangle=\textrm{e}^{-s\Phi(\lambda)}$, where $\Phi(\lambda)= \lambda^\alpha$ for $0<\alpha<1$ \cite{BauleFriedrich:2005} and $\Phi(\lambda)= \tau_0/(\alpha-1)\lambda-\tau_0|\Gamma(1-\alpha)| \lambda^\alpha$ for $1<\alpha<2$ \cite{WangChenDeng:2019}.  The two-point PDF of the subordinator $t(s)$ can be expressed as
\begin{equation}\label{g}
  g(t_1,s_1;t_2,s_2)= \langle \delta(t_1-t(s_1)) \delta(t_2-t(s_2)) \rangle.
\end{equation}
By virtue of the stationary and independent increments of subordinator $t(s)$, this two-point PDF in Laplace space ($t_1 \rightarrow \lambda_1, t_2\rightarrow\lambda_2$) is given by \cite{BauleFriedrich:2005}
\begin{eqnarray*}
    g(\lambda_1,s_1;\lambda_2,s_2)
    &=\Theta(s_2-s_1)\,{\rm e}^{-(s_2-s_1)\Phi(\lambda_2)}\,{\rm e}^{-s_1\Phi(\lambda_1+\lambda_2)}  \\
    &~+\Theta(s_1-s_2)\,{\rm e}^{-(s_1-s_2)\Phi(\lambda_1)}\,{\rm e}^{-s_2\Phi(\lambda_1+\lambda_2)},
\end{eqnarray*}
where $\Theta(x)$ is the Heaviside step function: $\Theta(x)=1$ for $x>0$, $\Theta(x)=0$ for $x<0$, and $\Theta(x=0)=1/2$.

The inverse $\alpha$-dependent subordinator $s(t):=\inf_{s>0}\{s:t(s)>t\}$ is defined as the first-passage time of the subordinator $t(s)$ with the two-point PDF
\begin{equation}
\begin{split}
    h(s_1,t_1;s_2,t_2)= \langle \delta(s_1-s(t_1)) \delta(s_2-s(t_2)) \rangle.
\end{split}
\end{equation}
The specific expression of $h(\cdot)$ can be obtained by taking the partial derivatives toward $s_1,s_2$ in the following equation \cite{BauleFriedrich:2005,WangChenDeng:2019}:
\begin{equation}\label{Thea}
\begin{split}
& \langle \Theta(s_2-s(t_2))\Theta(s_1-s(t_1))\rangle
    = 1-\langle\Theta(t_2-t(s_2))\rangle  \\
   &~~~ -\langle\Theta(t_1-t(s_1))\rangle
 +\langle\Theta(t_2-t(s_2))\Theta(t_1-t(s_1))\rangle,
\end{split}
\end{equation}
which links the two-point PDF of inverse subordinator to the corresponding subordinator. Then performing the Laplace transform ($t_1\rightarrow\lambda_1,t_2\rightarrow\lambda_2$), we obtain the two-point PDF $h(\cdot)$ of the inverse subordinator $s(t)$ in Laplace space as \cite{WangChenDeng:2019}
\begin{equation}\label{h2}
\begin{split}
&h(s_1,\lambda_1;s_2,\lambda_2)  \\
    &=\frac{\partial}{\partial s_1} \frac{\partial}{\partial s_2} \frac{1}{\lambda_1\lambda_2}\,g(\lambda_1,s_1;\lambda_2,s_2) \\
    &=\delta(s_2-s_1)\frac{\Phi(\lambda_1)+\Phi(\lambda_2)-\Phi(\lambda_1+\lambda_2)}{\lambda_1\lambda_2}\,{\rm e}^{-s_1\Phi(\lambda_1+\lambda_2)} \\
    &~~~+\Theta(s_2-s_1)\frac{\Phi(\lambda_2)(\Phi(\lambda_1+\lambda_2)-\Phi(\lambda_2))}{\lambda_1\lambda_2} \\
    &~~~\times{\rm e}^{-s_1\Phi(\lambda_1+\lambda_2)}{\rm e}^{-(s_2-s_1)\Phi(\lambda_2)}   \\
    &~~~+\Theta(s_1-s_2)\frac{\Phi(\lambda_1)(\Phi(\lambda_1+\lambda_2)-\Phi(\lambda_1))}{\lambda_1\lambda_2} \\
    &~~~\times{\rm e}^{-s_2\Phi(\lambda_1+\lambda_2)}{\rm e}^{-(s_1-s_2)\Phi(\lambda_1)}.
\end{split}
\end{equation}
The normalization of $h(s_2,\lambda_2;s_1,\lambda_1)$ can be verified through the equality $\int_0^\infty \int_0^\infty  h(s_2,\lambda_2;s_1,\lambda_1)ds_1 ds_2=(\lambda_1\lambda_2)^{-1}$.
Especially, the first term containing $\delta(s_2-s_1)$ in Eq. \eqref{h2} contributes to the part that no renewal happens between time $t_1$ and $t_2$. More precisely, let $p_0(t_1,t_2)$ be the probability of no renewal happens between time $t_1$ and $t_2$, and $\psi(\lambda)$ the Laplace transform of the waiting time distribution between two consecutive renewals. Then the double Laplace transform of $p_0(t_1,t_2)$ is \cite{GodrecheLuck:2001,WangChenDeng:2019-2}
\begin{equation}\label{p0lambda}
\begin{split}
  p_{0}(\lambda_1,\lambda_2)
  &\simeq  \frac{1+\psi(\lambda_1+\lambda_2)-\psi(\lambda_1)-\psi(\lambda_2)}
  {\lambda_1\lambda_2(1-\psi(\lambda_1+\lambda_2))}  \\
  &\simeq \frac{\Phi(\lambda_1)+\Phi(\lambda_2)-\Phi(\lambda_1+\lambda_2)}{\lambda_1\lambda_2\Phi(\lambda_1+\lambda_2)},
\end{split}
\end{equation}
where we have used the relation $\psi(\lambda)=g(\lambda,s)|_{s=1}\simeq1-\Phi(\lambda)$ for small $\lambda$ in the second line.
On the other hand, the result in Eq. \eqref{p0lambda} can be directly obtained by performing the double integrals with respect to $s_1$ and $s_2$ on the first term of the right-hand side of Eq. \eqref{h2}.
Correspondingly, another two terms in Eq. \eqref{h2} contribute to the probability that renewal happens between time $t_1$ and $t_2$ for $t_1<t_2$ and $t_1>t_2$, respectively. Explicitly presenting the relationship between the arguments of subordinator and renewal theory helps us better understand these two kinds of theories. Besides, it will be the method for us to simplify some derivations in the following sections.

Performing inverse Laplace transform of Eq. \eqref{p0lambda} yields \cite{FroembergBarkai:2013,WangChenDeng:2019}
\begin{equation}\label{p0t}
p_0(t_1,t_2)\simeq \left\{
\begin{array}{ll}
  \frac{\sin(\pi\alpha)}{\pi}B\left(\frac{t_1}{t_2};\alpha,1-\alpha \right),  & ~0<\alpha<1  \\[5pt]
  (t_2-t_1)^{1-\alpha}-t_2^{1-\alpha}, &~ 1<\alpha<2
\end{array}\right.
\end{equation}
for $t_1<t_2$. The results in Eq. \eqref{p0t} are the asymptotic forms for large difference between $t_1$ and $t_2$.
More detailed messages of the quantities of the stochastic processes are commonly determined by the multi-point PDFs of the inverse subordinator $s(t)$, especially the two-point one in Eq. \eqref{h2}. In the following sections, we will establish the Langevin picture to describe the L\'{e}vy walk in the presence of a constant force and evaluate some concerned statistical quantities through the two- and four-point PDFs of inverse subordinator.

\section{L\'{e}vy walk in a constant force field}\label{three}

The L\'{e}vy walk model with constant force $F$ can be described as a set of Langevin equations coupled with a subordinator as
\begin{equation}\label{LW_force}
\begin{split}
    \frac{d}{d t}x(t)&=v(t),\\
    \frac{d}{d s}v(s)&=-\gamma v(s) +F \eta(s)+\xi(s),\\
    \frac{d}{d s}t(s)&= \eta(s),
\end{split}
\end{equation}
where $\gamma$ is the friction coefficient, $\xi(s)$ is a Gaussian white noise with null mean value $\langle \xi(s)\rangle=0$ and $\langle \xi(s_1)\xi(s_2)\rangle=2D\delta(s_1-s_2)$. The L\'{e}vy noise $\eta(s)$, regarded as the formal derivative of the $\alpha$-dependent subordinator $t(s)$, is independent with the Gaussian white noise $\xi(s)$. The derivative of position $x$ with respect to physical time $t$ is velocity $v$ and the subordinator $t(s)$ is aimed to characterize the distribution of duration of each flight of L\'{e}vy walk. The initial position and velocity are assumed to be $x(0)=v(0)=0$. When $F=0$, the Langevin picture \eqref{LW_force} is coincided with the free-force case \cite{WangChenDeng:2019}.

The constant force $F$ multiplied by L\'{e}vy noise $\eta(s)$ is meant to affect the stochastic process for all physical time $t$ after making the subordination \cite{CairoliBaule:2015-2,ChenWangDeng:2019-2}. Otherwise, the force $F$ is only effective over operational time $s$, similar to the case that the external force only affects on the instant of jump in CTRW model, and it is invalid during the trap event with the constant $s$ \cite{CompteMetzlerCamacho:1997,Compte:1997}.

To be more intuitive, the Langevin equation for the velocity process in Eq. \eqref{LW_force} can be transformed to the one in physical time $t$ by using the identity $v(t):=v(s(t))$, i.e.,
\begin{equation}\label{LWeq}
\begin{split}
\frac{d}{d t}v(t)=-\gamma v(t) \frac{d}{d t}s(t)+F+\xi(s(t))\frac{d}{d t}s(t),
\end{split}
\end{equation}
where $\xi(s(t))$ can be seen as the time-changed Gaussian white noise \cite{CairoliBaule:2017}. Equation \eqref{LWeq} implies that the external force $F$ affects the velocity process for all physical time $t$. Especially when the inverse subordinator $s(t)$ remains a constant, this equation reduces to $dv(t)/dt=F$, characterizing the fact that the constant force provides an acceleration $F$ to the unidirectional motion of L\'{e}vy walk.

\begin{figure}[!htb]
\begin{minipage}{0.32\linewidth}
  \centerline{\includegraphics[scale=0.4]{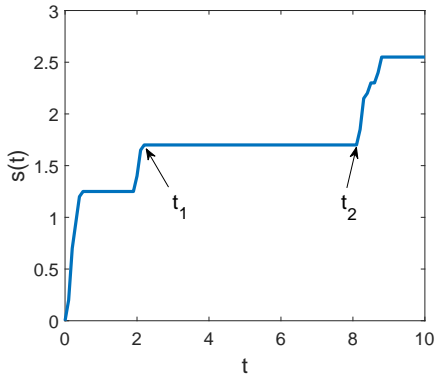}}
\end{minipage}
\caption{{(Color online)} Sample trajectory of inverse $\alpha$-dependent subordinator $s(t)$ with $\alpha=0.6$.}\label{st}
\end{figure}

On the other hand, the velocity process $v(s)$ can be solved from the second equation of Eq. \eqref{LW_force}, by using the Laplace transform technique, i.e.,
\begin{equation}\label{LWs}
\begin{split}
v(s)=F\int_0^s e^{-\gamma(s-s')}\eta(s')ds'+\int_0^s e^{-\gamma(s-s')}\xi(s')ds'.
\end{split}
\end{equation}
Equation \eqref{LWs} shows that the velocity $v(s)$ is contributed by two parts; one comes from the external force $F$ and another one from the random force $\xi(s)$ which corresponds to the free L\'{e}vy walk. Due to the independence of noises $\eta(s)$ and $\xi(s)$, the two terms in Eq. \eqref{LWs} are independent, and the subordinated process $v(t):=v(s(t))$ in physical time $t$ is given by
\begin{equation}\label{LWt}
\begin{split}
v(t)&=F\int_0^t e^{-\gamma(s(t)-s(t'))}dt'  \\
&~~~+\int_0^t e^{-\gamma(s(t)-s(t'))}\xi(s(t'))ds(t').
\end{split}
\end{equation}
See Appendix \ref{App1} for detailed derivations of Eqs. \eqref{LWeq} and \eqref{LWt}. Equations \eqref{LWs} and \eqref{LWt} are complementary for further quantitative calculations in the following sections. Here, the first term of velocity in Eq. \eqref{LWt} is resulted from the contribution of the constant force $F$. Instead of the naive expectation that this contribution is $Ft$ during time $t$, it is found to be not larger than $Ft$ due to the non-decreasing property of $s(t)$.
The expression of $v(t)$ in Eq. \eqref{LWt} means that the evolution of velocity process starts from time zero, and the force's contribution reaches the maximum $Ft$ only if $s(t')\equiv 0$ when $t'\in[0,t]$, which implies no renewal happens during the measurement time $t$.
Once the renewal happens, the system will lose some of the extra momentums provided by the constant force, and then the net increased velocity is less than $Ft$.
In particular, let us focus on the value of velocity between any two successive renewal points.
As the graph of one sample of inverse subordinator $s(t)$ in Fig. \ref{st} shows, if no renewal happens between the renewal points $t_1$ and $t_2$, the $s(t)$ will remain a constant for $t\in[t_1,t_2]$. By using this property of $s(t)$, we obtain the increment of velocity process in Eq. \eqref{LWt} between two successive renewal points
\begin{equation}\label{v12}
\begin{split}
v(t_2)-v(t_1)&=F\int_{t_1}^{t_2} e^{-\gamma(s(t_1)-s(t'))}dt'\\
             &=F(t_2-t_1),
\end{split}
\end{equation}
where the second term in Eq. \eqref{LWt} vanishes for constant $s(t)$. Equation \eqref{v12} implies that the acceleration in the no-renewal period is constant $F$.

An alternative description of L\'{e}vy walk in a constant force that we have to mention is the collision model in Ref. \cite{BarkaiFleurov:1998}. It says that the change of the velocity of a free test particle is due to the collision with the surrounding bath particles. If a constant force $F$ is added, the acceleration is $F$ between two successive collisions. However, some of the added momentum in the history will be transmitted to the surrounding bath particles when the collision happens according to the energy and momentum conservation laws.

Now, the coupled Langevin equation in Eq. \eqref{LW_force} is a continuous limit of the collision model in some sense. The Gaussian white noise $\xi(s)$ in operation time $s$ describes the random force coming from the collision with the surrounding small bath particles, and the subordinator $t(s)$ corresponds to the random time distributions between two successive collisions by taking $s=1$. Besides the intuitive analyses, we will show that the coupled Langevin system \eqref{LW_force} and the collision model are quantitatively consistent in some important quantities---the means of velocity $v(t)$ and displacement $x(t)$.
 Furthermore, based on the
explicit expressions of velocity process in Eqs. \eqref{LWs} and \eqref{LWt}, the correlation function of velocity, EAMSD and TAMSD, which have not been obtained before, can also be evaluated.


\section{First moment}\label{four}
This section is aimed to derive the mean value of $v(t)$ and $x(t)$ in Eq. \eqref{LW_force} by using the velocity process in Eqs. \eqref{LWs} and \eqref{LWt}, together with the subordination method \cite{BauleFriedrich:2005,ChenWangDeng:2019-2,ChenWangDeng:2018-2,WangChenDeng:2019}.
The mean value of velocity process in physical time can be obtained by directly taking the average on Eq. \eqref{LWt},
\begin{equation}\label{v}
\begin{split}
\langle v(t)\rangle=F\int_0^t \langle e^{-\gamma(s(t)-s(t'))}\rangle dt'.
\end{split}
\end{equation}
The ensemble average of the second term in Eq. \eqref{LWt} vanishes due to the zero mean value of $\xi(s)$.  It is interesting to find that the first moment of the velocity process in Eq. \eqref{v} depends on the two-point PDF of the inverse subordinator.
By using the two-point PDF $h(s,t;s',t')$ in \eqref{h2} and the technique of Laplace transform ($t\rightarrow \lambda,t'\rightarrow \lambda'$), the integrand in Eq. \eqref{v} can be obtained, for small $\lambda$ and $\lambda'$, as
\begin{equation}\label{FMintegral}
\begin{split}
\mathcal{L} [\langle e^{-\gamma s(t)}e^{\gamma s(t')}\rangle]
&=\int_0^\infty \int_0^\infty e^{-\gamma (s-s')}h(s,\lambda;s',\lambda')ds'ds\\
&\simeq \frac{\Phi(\lambda)+\Phi(\lambda')-\Phi(\lambda+\lambda')}{\lambda\lambda'\Phi(\lambda+\lambda')},
\end{split}
\end{equation}
which is the same as $ p_0(\lambda',\lambda)$ in Eq. \eqref{p0lambda}.
The rationality of small $\lambda$ and $\lambda'$ is that we pay attention to the large-$t$ behavior of this system and the integrand of Eq. \eqref{v} is dominated by the part with large $t'$. Equation \eqref{FMintegral} implies that the integrand of Eq. \eqref{v} is asymptotically equal to $p_0(t',t)$ with $t'<t$.
This result indicates that the first term of two-point PDF $h(s_1,t_1;s_2,t_2)$ in Eq. \eqref{h2} containing $\delta$-function  plays a leading role when integrated together with an exponential kernel $e^{-\gamma(s_1-s_2)}$ in Eq. \eqref{FMintegral}.

Substituting the expression $p_0(t',t)$ in Eq. \eqref{p0t} and calculating the integral in Eq. \eqref{v} yield
\begin{equation}\label{mean-v}
\begin{split}
\langle v(t)\rangle  \simeq \left\{
    \begin{array}{ll}
    F(1-\alpha)t, &~~ 0<\alpha<1, \\[4pt]
    F\frac{\alpha-1}{2-\alpha}t^{2-\alpha}, &~~ 1<\alpha<2.
\end{array}
  \right.
\end{split}
\end{equation}
The mean value of $x(t)$ affected by the constant force $F$ is the integral of $\langle v(t)\rangle$, i.e.,
\begin{equation}\label{x}
\begin{split}
\langle x(t)\rangle_F \simeq \left\{
    \begin{array}{ll}
       F\frac{(1-\alpha)}{2}t^2, & 0<\alpha<1, \\[4pt]
      F\frac{(\alpha-1)}{(2-\alpha)(3-\alpha)}t^{3-\alpha}, & 1<\alpha<2.
\end{array}
  \right.
\end{split}
\end{equation}
Considering the EAMSD $\langle x^2(t)\rangle_0$ of free L\'{e}vy walk \cite{FroembergBarkai:2013,WangChenDeng:2019} and the first moment in presence of the constant force $F$ in Eq. \eqref{x}, the generalized Einstein relation \cite{BouchaudGeorges:1990,BarkaiMetzlerKlafter:2000,MetzlerKlafter:2000} is satisfied by L\'{e}vy walk \cite{FroembergBarkai:2013-3}
\begin{equation}
  \langle x(t)\rangle_F = \frac{\langle x^2(t)\rangle_0}{2k_B\mathcal{T}}F,
\end{equation}
where the effective kinetic temperature $k_B\mathcal{T}$ is equal to $D/\gamma$ for the Langevin system in Eq. \eqref{LW_force}. However, we find that the generalized Einstein relation for time averages does not hold, i.e.,
\begin{equation}
\begin{split}
\langle \overline{\delta^1(\Delta)}\rangle_F \neq \frac{\langle\overline{\delta^2(\Delta)} \rangle_0}{2k_B \mathcal{T}}F
\end{split}
\end{equation}
with $\langle\overline{\delta^1(\Delta)}\rangle_F=\frac{1}{T-\Delta}\int_0^{T-\Delta}\langle x(t+\Delta)-x(t)\rangle_F dt$ and  $\langle\overline{\delta^2(\Delta)} \rangle_0$ being the TAMSD of free L\'{e}vy walk. Actually, for L\'{e}vy walk, there is \cite{HeBurovMetzlerBarkai:2008,FroembergBarkai:2013-3}
\begin{equation}
\begin{split}
\frac{\langle \overline{\delta^1(\Delta)}\rangle_F}{\langle\overline{\delta^2(\Delta)} \rangle_0}=\frac{F|\alpha-1|}{2k_B\mathcal{T}}\left(\frac{T}{\Delta}\right)^{\tilde{\alpha}-1},
\end{split}
\end{equation}
which is consistent with the result in Ref. \cite{FroembergBarkai:2013-3}. Here, $\tilde{\alpha}=2$ when $0<\alpha<1$ and $\tilde{\alpha}=3-\alpha$ when $1<\alpha<2$.

\section{velocity correlation function}\label{five}
Both the EAMSD and TAMSD of $x(t)$ depend on the velocity correlation function $\langle v(t_1)v(t_2)\rangle$, which is much more difficult to calculate than the mean value in Section \ref{four}. Considering the independence of the two terms in Eq. \eqref{LWs}, the velocity correlation function only consists of two parts:
\begin{equation}\label{TwoParts}
\begin{split}
\langle v(t_1)v(t_2)\rangle=\langle v(t_1)v(t_2)\rangle_1+\langle v(t_1)v(t_2)\rangle_2.
\end{split}
\end{equation}
The first term comes from the $F$-dependent part in Eq. \eqref{LWt},
\begin{equation}\label{vv}
\begin{split}
\langle v(t_1)v(t_2)\rangle_1= F^2\int_0^{t_1} \int_0^{t_2}\langle W(t_1,t_1',t_2,t_2') \rangle dt_2'dt_1'.
\end{split}
\end{equation}
We use the symbol
\begin{equation}\label{SymbolW}
  W(t_1,t_1',t_2,t_2')=e^{-\gamma s(t_1)}e^{\gamma s(t_1')}e^{-\gamma s(t_2)}e^{\gamma s(t_2')}
\end{equation}
to denote the double exponential kernel, the ensemble-average of which depends on the four-point PDF of inverse subordinator $s(t)$.
As for the second term in Eq. \eqref{TwoParts}, it is the same as the free-force case and is more convenient to be derived in operation time from Eq. \eqref{LWs} and then apply the subordination method, i.e.,
\begin{equation}\label{vv2}
\begin{split}
& \langle v(t_1)v(t_2)\rangle_2
\\
& =\int_0^\infty\!\!\!\!\int_0^\infty \langle v(s_1)v(s_2)\rangle_2 h(s_1, t_1; s_2, t_2)ds_1ds_2,\\
\end{split}
\end{equation}
where the velocity correlation function $\langle v(s_1)v(s_2)\rangle_2$ in operation time can be obtained through the second part of $v(s)$ in Eq. \eqref{LWs}
\begin{equation}
\begin{split}
\langle v(s_1)v(s_2)\rangle_2= \frac{D}{\gamma} (e^{-\gamma |s_1- s_2|}-e^{-\gamma(s_1+s_2)}).
\end{split}
\end{equation}
Then by use of the two-point PDF of the inverse subordinator in Eq. \eqref{h2}, we obtain
\begin{equation}\label{VV2}
\langle v(t_1)v(t_2)\rangle_2 \simeq \frac{D}{\gamma}p_0(t_1,t_2)
\end{equation}
for large $t_1$, $t_2$ and $t_1<t_2$, which is the same as the velocity correlation function of free L\'{e}vy walk in CTRW framework when the constant $D/\gamma=v_0^2$ \cite{FroembergBarkai:2013,WangChenDeng:2019}.

The remaining question is how to evaluate $\langle W(t_1,t_1',t_2,t_2') \rangle$ in the first part of velocity correlation function in Eq. \eqref{TwoParts}, which looks quite complex since it depends on the four-point PDF of inverse subordinator $s(t)$. However, based on the
analyses of Eq. \eqref{FMintegral}, we find that the leading term of four-point PDF should contain $\delta(t_1-t_1')\delta(t_2-t_2')$ due to the exponential kernel $W$, and the value of $\langle W(t_1,t_1',t_2,t_2') \rangle$ might be closely related to $p_0(t_1',t_1)$ and $p_0(t_2',t_2)$.

More precisely, the value of $\langle W(t_1,t_1',t_2,t_2') \rangle$ depends on the magnitude relation between the four time points $t_1,t_1',t_2,t_2'$. Since $t_1'<t_1$ and $t_2'<t_2$ in Eq. \eqref{vv} and it is assumed that $t_1<t_2$, there are totally three different cases for different range of $t_2'$. 
For the first two cases, there are
\begin{equation}\label{W12}
\begin{split}
\langle W(t_1,t_1',t_2,t_2') \rangle \simeq \left\{
\begin{array}{ll}
     p_0(t_2',t_2), & t_2'<t_1'<t_1<t_2, \\[4pt]
     p_0(t_1',t_2), & t_1'<t_2'<t_1<t_2.
\end{array}\right.
\end{split}
\end{equation}
The situation becomes more complex for the case $t_1'<t_1<t_2'<t_2$. In this case, $\langle W(t_1,t_1',t_2,t_2') \rangle$ consists of two parts; one is $p_0(t_1',t_2)$ denoting no renewal happens throughout $[t_1',t_2]$ and another one denoting some renewals happen in the middle interval $[t_1,t_2']$. The former one is easy to be calculated as those in Eq. \eqref{W12},  while the latter one depends on the four-point PDF of inverse subordinator $s(t)$ and it should be contributed from the term containing $\delta(t_2-t_2')\Theta(t_2'-t_1)\delta(t_1-t_1')$ in the four-point PDF of inverse subordinator. With some detailed derivations in Appendix \ref{App2}, we obtain the result for $t_1'<t_1<t_2'<t_2$,
\begin{equation}\label{W3}
\begin{split}
\langle W(t_1,t_1',t_2,t_2') \rangle \simeq \left\{
\begin{array}{ll}
     p_0(t_1',t_1)p_0(t_2',t_2), & 0<\alpha<1, \\[4pt]
     p_0(t_1',t_2), & 1<\alpha<2.
\end{array}\right.
\end{split}
\end{equation}

For the case $t_1<t_2$, splitting the integral Eq. \eqref{vv} into three parts according to the range of $t_2'$ as
\begin{equation*}
\begin{split}
    \int_0^{t_1}dt_1' \int_0^{t_2}dt_2' =& \int_0^{t_1}dt_1' \int_0^{t_1'}dt_2' +\int_0^{t_1}dt_1' \int_{t_1'}^{t_1}dt_2' \\
    &+\int_0^{t_1}dt_1' \int_{t_1}^{t_2}dt_2'
\end{split}
\end{equation*}
and using the corresponding forms of $\langle W(t_1,t_1',t_2,t_2') \rangle$, we obtain, for large $t_1$ and $t_2$,
\begin{widetext}
\begin{equation}\label{VV}
\langle v(t_1)v(t_2)\rangle_1 \simeq \left\{
\begin{array}{ll}
      \frac{F^2\sin(\pi\alpha)}{\pi}\left[ t_2^2B\left(\frac{t_1}{t_2};\alpha+2,1-\alpha\right) + \alpha t_1^2B\left(\frac{t_1}{t_2};\alpha,1-\alpha\right) \right.  \\[5pt]
        ~~~~~~~~~~~~~~\left. -(1+\alpha)t_1t_2B\left(\frac{t_1}{t_2};\alpha+1,1-\alpha\right) \right]  + F^2 (1-\alpha)^2\,t_1t_2, &~~0<\alpha<1, \\[8pt]
      \frac{F^2(\alpha-1)}{2-\alpha}\left(t_1t_2^{2-\alpha}-\frac{1}{3-\alpha}t_2^{3-\alpha}+\frac{1}{3-\alpha}(t_2-t_1)^{3-\alpha}\right), & ~~1<\alpha<2. \\
\end{array} \right.
\end{equation}
\end{widetext}

\section{EAMSD and TAMSD}\label{six}
Once the velocity correlation function is obtained, we can rewrite the summation of Eqs. \eqref{VV2} and \eqref{VV} as the scaling form with two separate parts 
\begin{equation}\label{ScalFormV}
  \langle v(t_1)v(t_2)\rangle \simeq C_1t^{\nu_1-2}\phi_1\left(\frac{\tau}{t}\right)+C_2t^{\nu_2-2}\phi_2\left(\frac{\tau}{t}\right)
\end{equation}
for large $t$ and $\tau$. See Appendix \ref{App3} for the expressions of $C_{i}$, $\nu_{i}$, $\phi_{i}$ ($i=1,2$). We use two different scaling forms since the velocity correlation function contains two independent parts in Eq. \eqref{TwoParts}.
The dependence of EAMSD and TAMSD on the scaled velocity correlation function has been revealed through the generalized Green-Kubo formula \cite{DechantLutzKesslerBarkai:2014,MeyerBarkaiKantz:2017}.

\begin{figure*}[!htb]
\begin{minipage}{0.32\linewidth}
  \centerline{\includegraphics[scale=0.4]{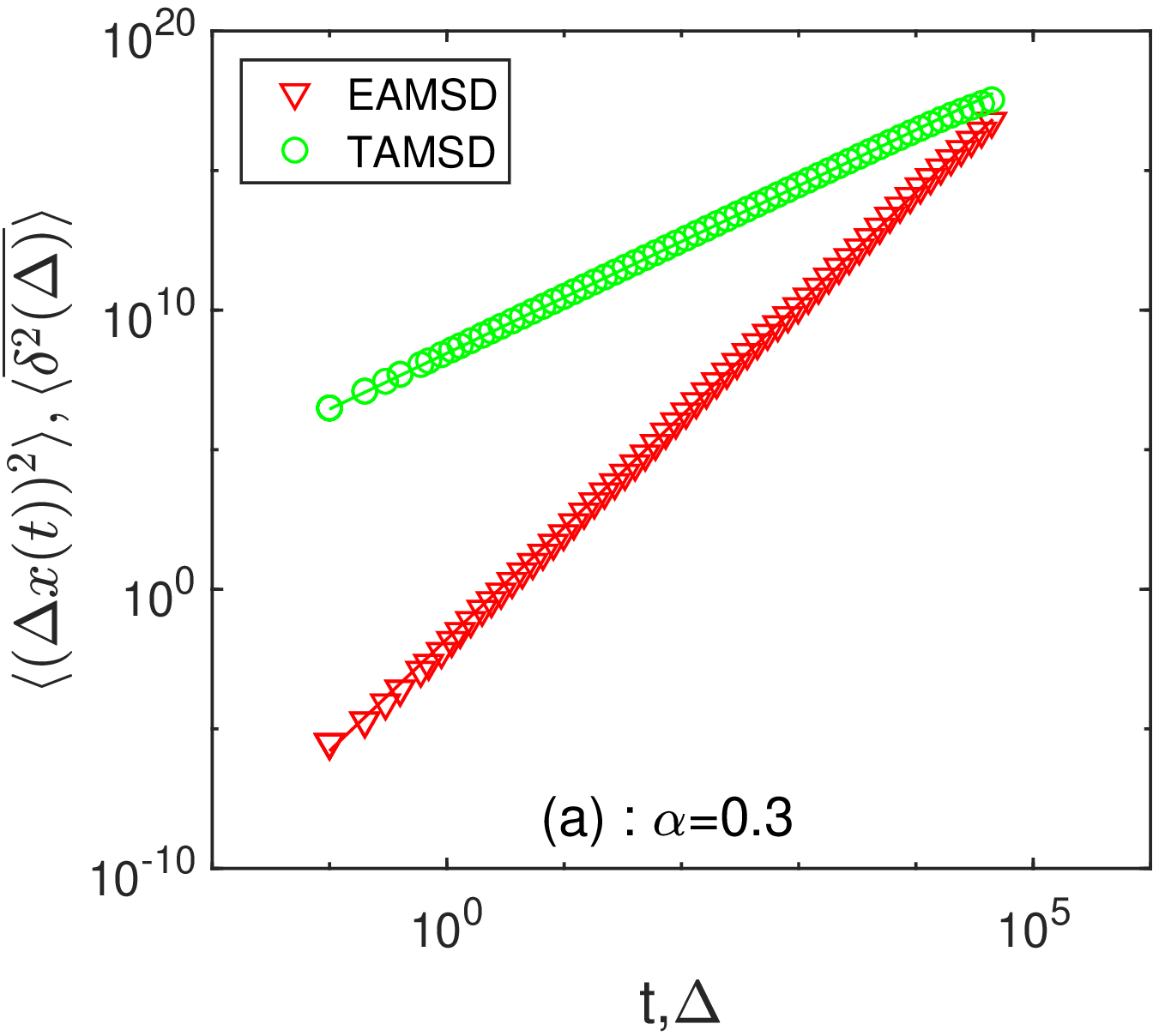}}
\end{minipage}
\hfill
\begin{minipage}{0.32\linewidth}
  \centerline{\includegraphics[scale=0.4]{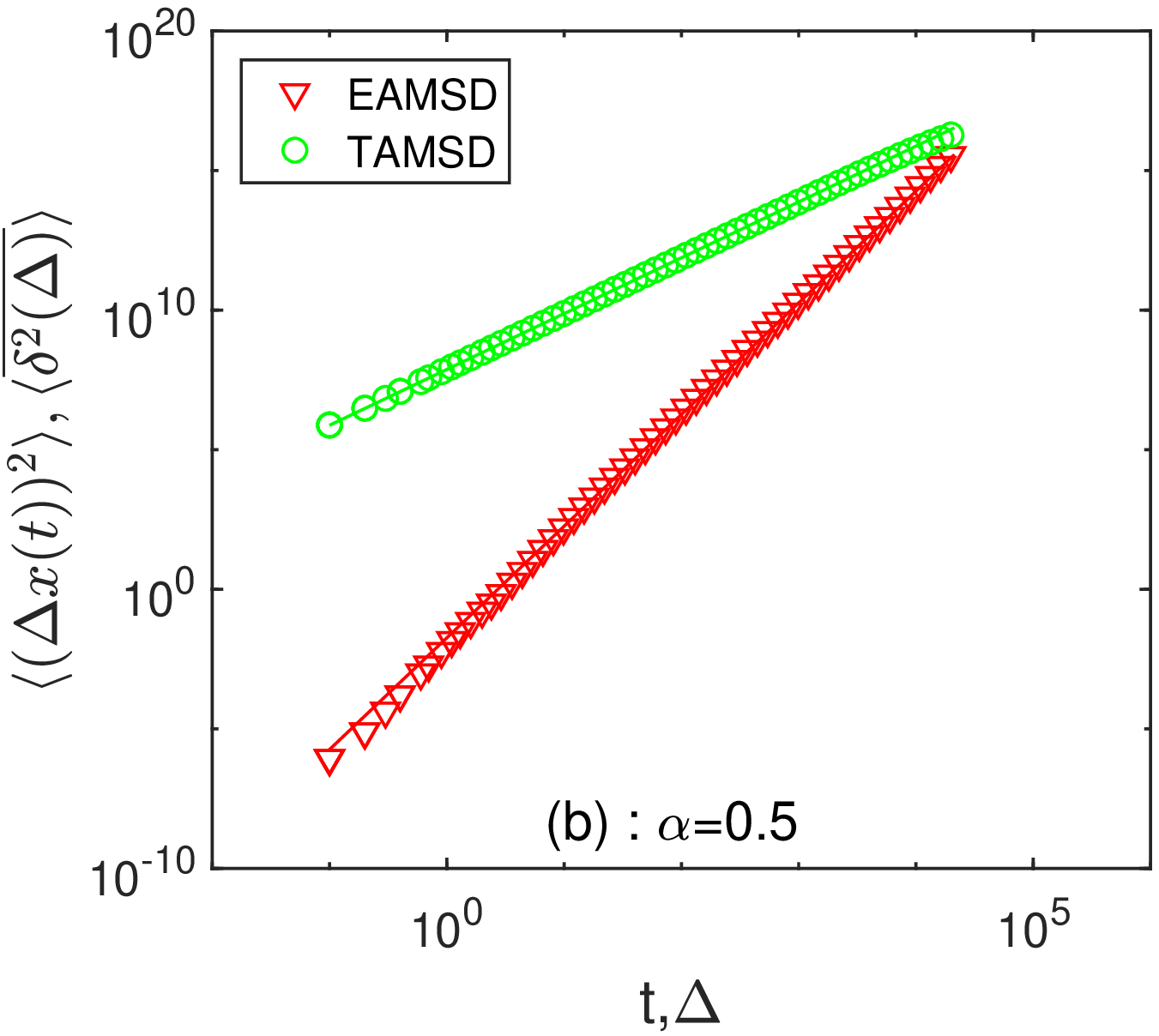}}
\end{minipage}
\hfill
\begin{minipage}{0.32\linewidth}
  \centerline{\includegraphics[scale=0.4]{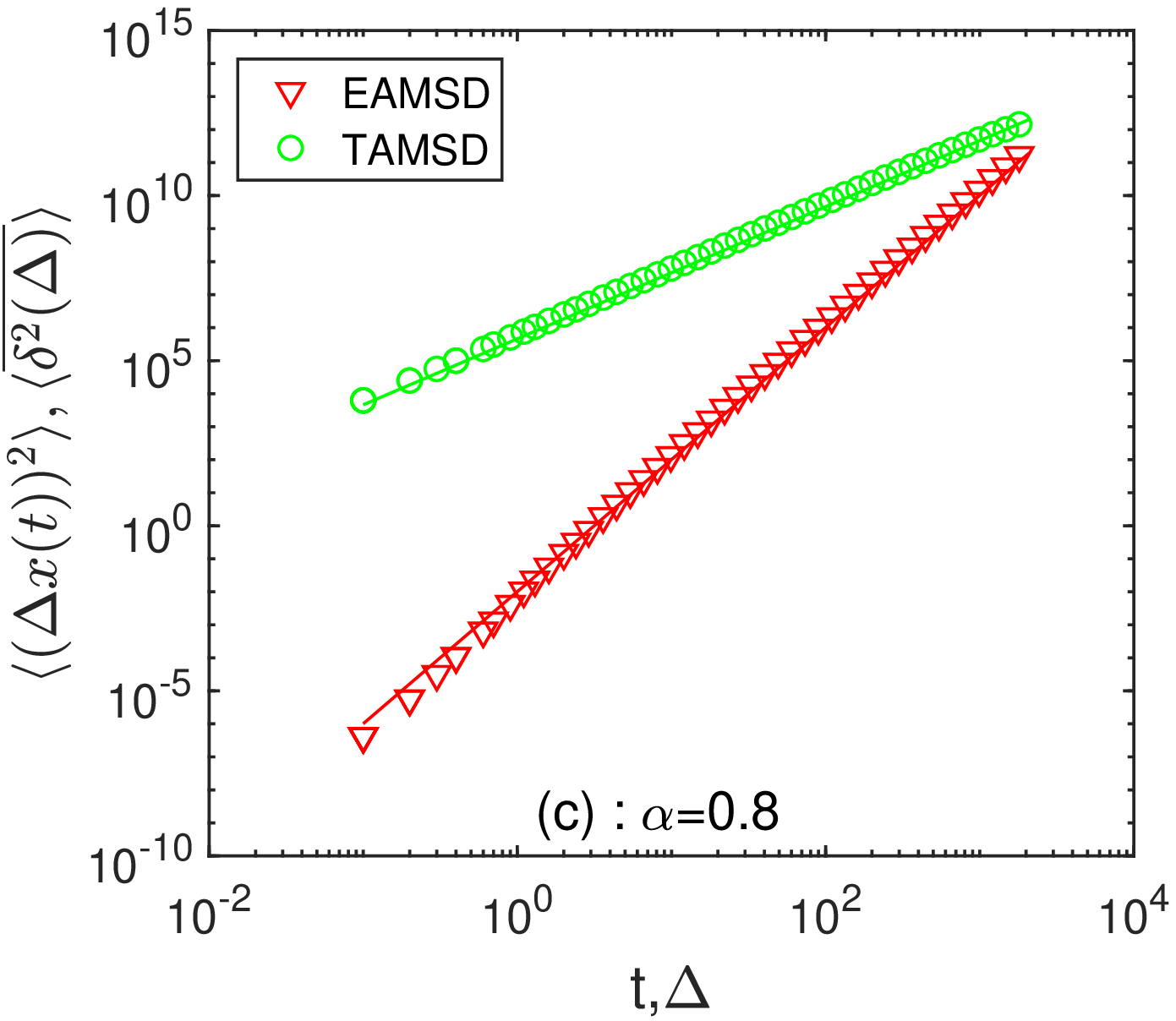}}
\end{minipage}
\vfill
\begin{minipage}{0.32\linewidth}
  \centerline{\includegraphics[scale=0.4]{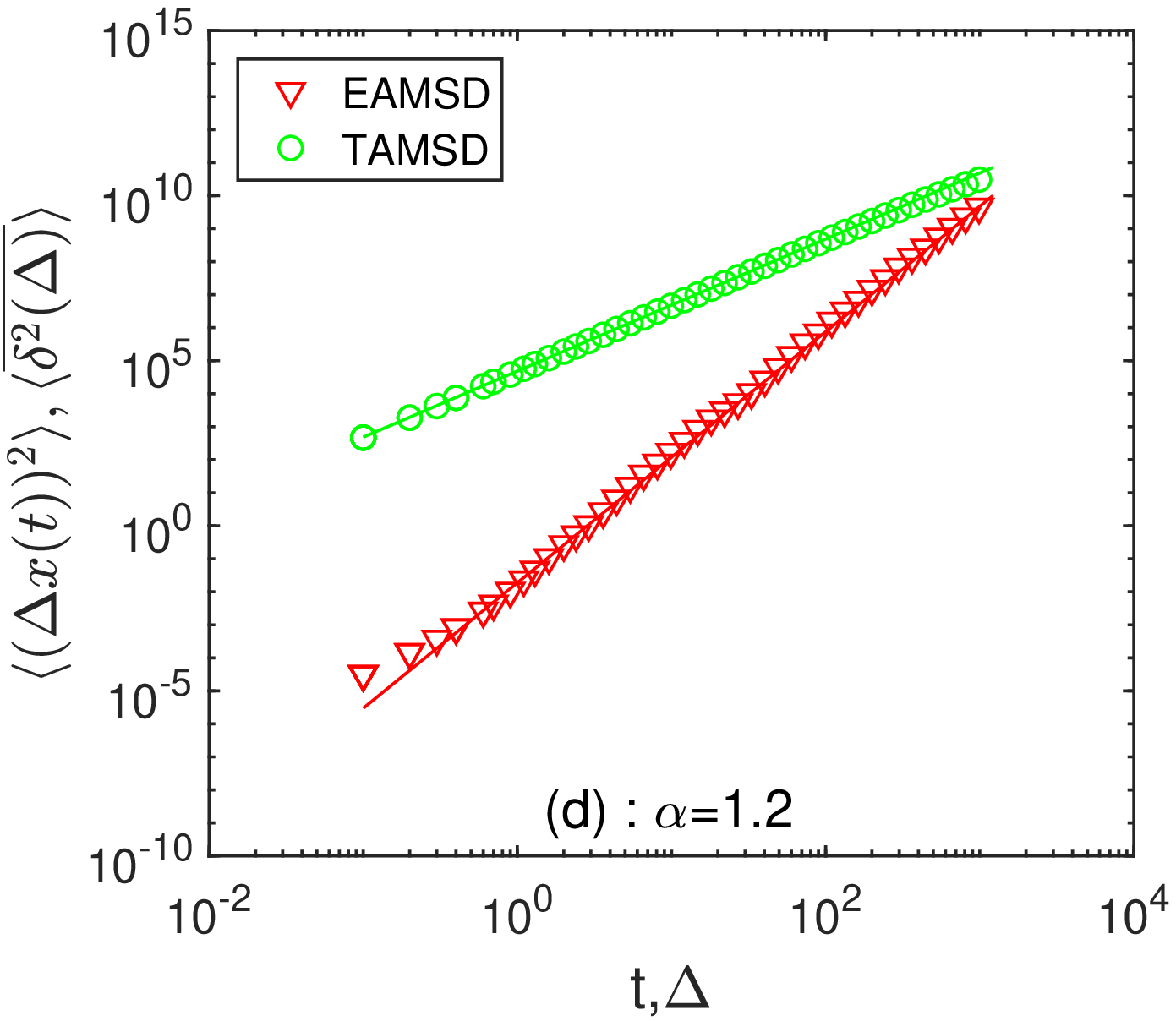}}
\end{minipage}
\hfill
\begin{minipage}{0.32\linewidth}
  \centerline{\includegraphics[scale=0.4]{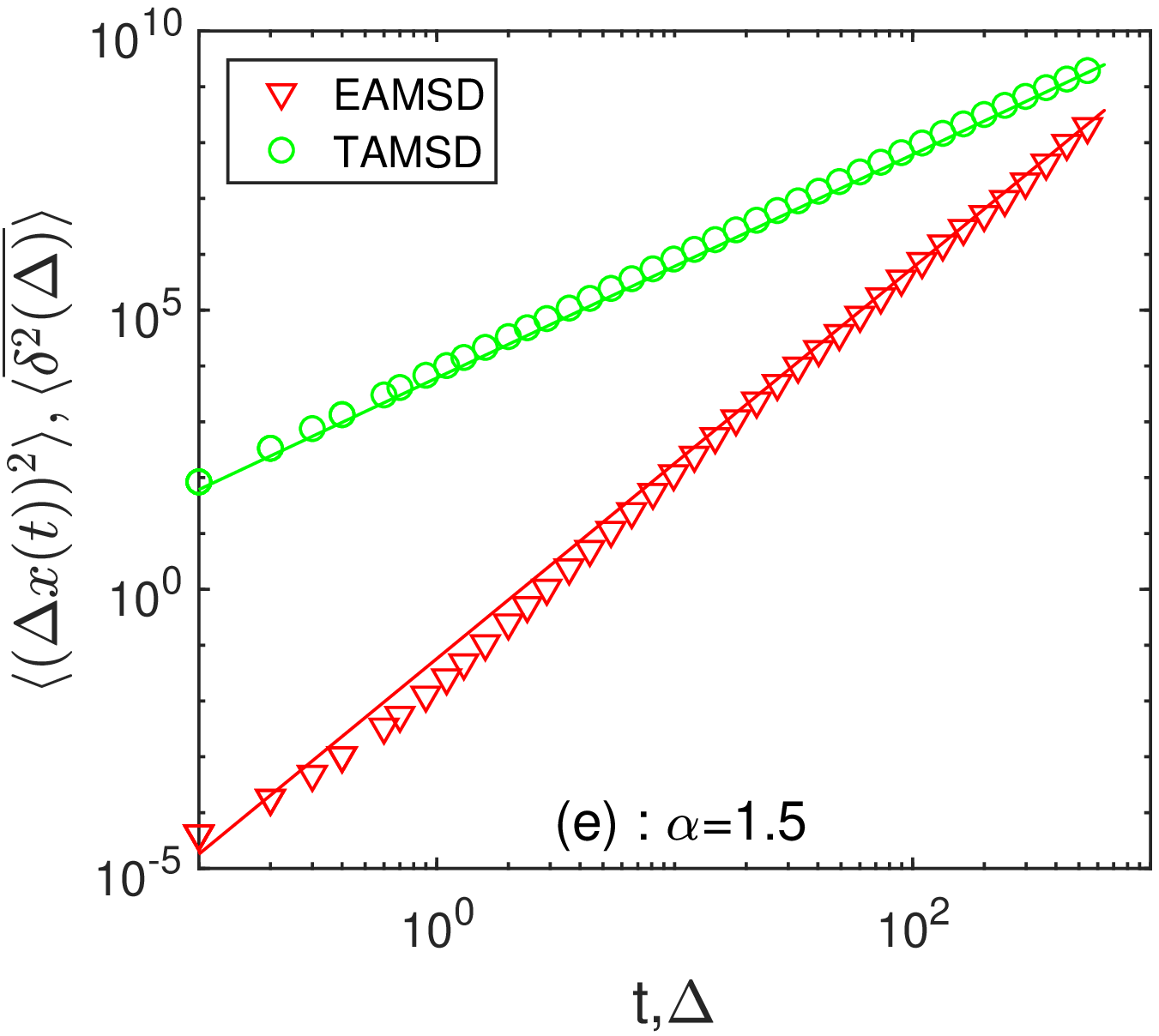}}
\end{minipage}
\hfill
\begin{minipage}{0.32\linewidth}
  \centerline{\includegraphics[scale=0.4]{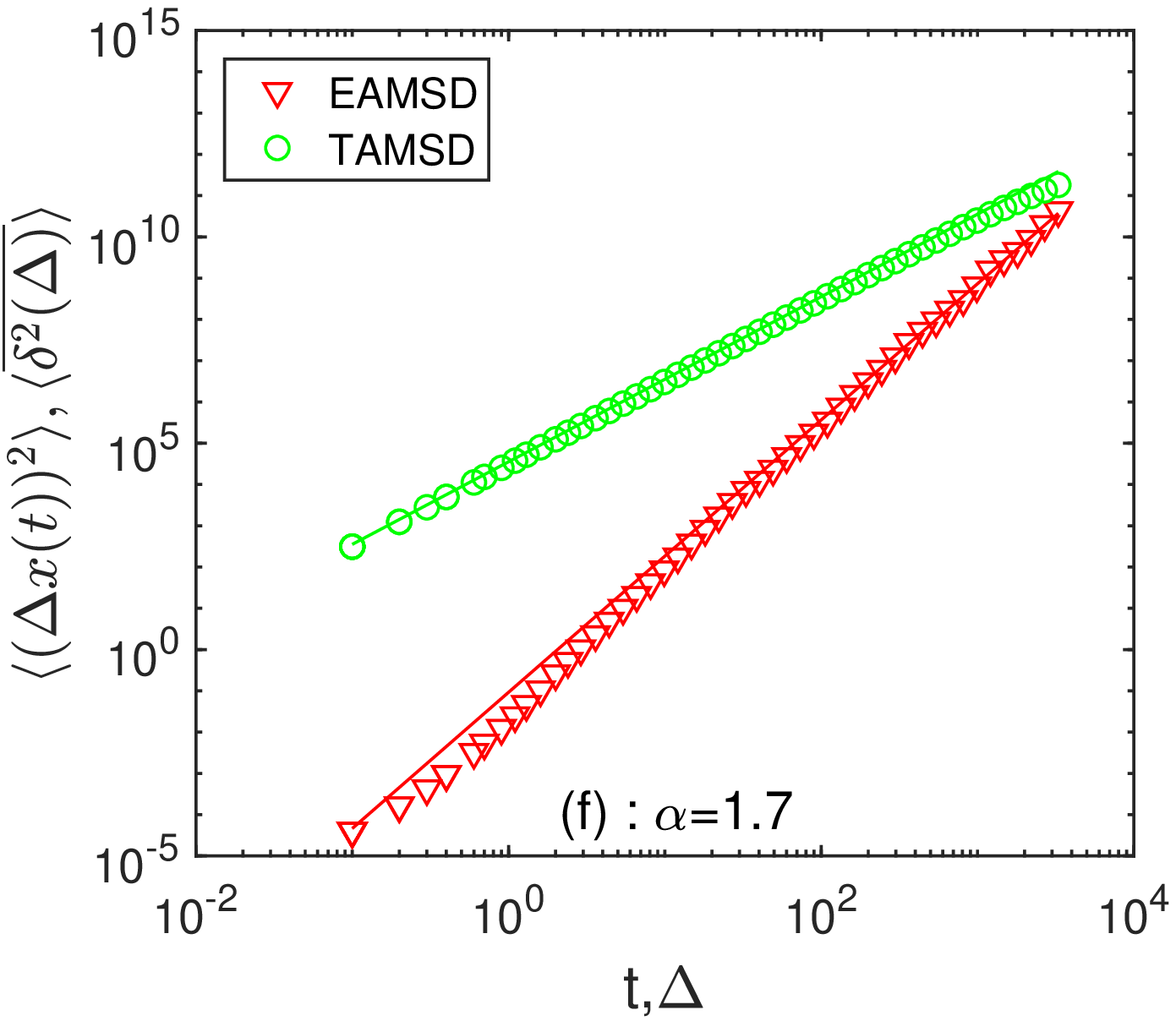}}
\end{minipage}
\caption{{(Color online)} EAMSD and TAMSD of the process described by the Langevin equation \eqref{LW_force} for different $\alpha$. The parameters are taken as $F=1$, $\gamma=2$, and $D=1$.
Red triangle-markers and green circle-markers are the simulation results for EAMSD and TAMSD,  which coincide with the theoretical results of Eqs. \eqref{EA} and \eqref{TA} being represented by solid lines. }\label{fig2}
\end{figure*}

Here what we should pay attention to is to use the generalized Green-Kubo formula to handle the two parts in Eq. \eqref{ScalFormV} individually and then combine them together. After the lengthy calculations in Appendix \ref{App3}, we obtain the EAMSD of L\'{e}vy walk in the presence of a constant force as
\begin{equation}\label{EA}
\begin{split}
\langle(\Delta x(t))^2\rangle&=\langle x^2(t)\rangle-\langle x(t)\rangle^2  \\[4pt]
&\simeq \left\{
\begin{array}{ll}
      \frac{F^2\alpha(1-\alpha)(2-\alpha)(3+\alpha)}{72}t^4, & 0<\alpha<1, \\[4pt]
      \frac{F^2 (\alpha-1)}{(4-\alpha)(5-\alpha)}t^{5-\alpha}, & 1<\alpha<2, \\[4pt]
\end{array}\right.
\end{split}
\end{equation}
both of which present the super-ballistic diffusion behavior resulting from the acceleration. For $0<\alpha<1$, the frequency of the collision is too low so that the ensemble-averaged acceleration keeps proportional to $F$ in Eq. \eqref{mean-v}. While for $1<\alpha<2$, the collision happens more frequently then the ensemble-averaged acceleration decays as $t^{1-\alpha}$ in Eq. \eqref{mean-v} and the diffusion behavior is slower than $t^4$.
The corresponding TAMSD is
\begin{equation}\label{TA}
\begin{split}
\langle \overline{\delta^2(\Delta)}\rangle
&\simeq
                     \left\{
    \begin{array}{ll}
    \frac{F^2(1-\alpha)\alpha}{6}T^2\Delta^2, & 0<\alpha<1, \\[4pt]
    \frac{F^2 (\alpha-1)}{(3-\alpha)(4-\alpha)}T^{3-\alpha}\Delta^2, & 1<\alpha<2, \\[4pt]
\end{array}
  \right.
\end{split}
\end{equation}
for large $T$ and $\Delta\ll T$. Different from the free L\'{e}vy walk, both of the TAMSDs here are increasing as $\propto\Delta^2$ whatever $0<\alpha<1$ or $1<\alpha<2$. This system is weakly nonergodic since the EAMSD and TAMSD are different in the exponents of $t$ and $\Delta$. In Fig. \ref{fig2}, the simulations of EAMSDs and TAMSDs for different $\alpha$ agree with the theoretical results very well.

\section{Summary}\label{seven}
L\'{e}vy walk is a typical and practical model and it has various applications in the natural world.
In this paper, we mainly investigate the anomalous and nonergodic behavior of the L\'{e}vy walk under the effects of a constant force in physical times. The Langevin picture of such a model has been established and compared with a previously proposed collision model in \cite{BarkaiFleurov:1998}.
The consistency between the two models has been illustrated/verified through the intuitive descriptions of the models and the quantitative comparison of the first moment of velocity process.

Compared with the collision model, one of the advantages of Langevin equation is that the velocity correlation function can be conveniently obtained by using the techniques of subordination. As for the method of subordination, we get some new findings in this paper. The first one is the relationship between the PDF of inverse subordinator and the PDF of the number of renewals. The second one is that by adding the term $F\eta(s)$ in the Langevin equation on operation time $s$ (in Eq. \eqref{LW_force}), we obtain an effective force being active for all physical times. With some changes of the formulation of the Langevin equation, we clearly show the effects of the external force on the Langevin system and connect it with the collision model. The third one is an interesting and rare phenomenon that the first moments of the velocity and displacement depend on the two-point PDF of the inverse subordinator. More complicated, the velocity correlation function depends on the four-point PDF of the inverse subordinator. By using the relationship between the method of subordinator and the renewal theory, we simplify the derivations and obtain the velocity correlation function.

After getting the velocity correlation function, we can obtain the EAMSD and TAMSD by using the generalized Green-Kubo formula. The super-ballistic diffusion and weakly non-ergodic behavior are observed for this Langevin system. Apart from the interesting phenomena for L\'{e}vy walk in the presence of a constant force, we are also interested in the L\'{e}vy walk under the effects of other kinds of external forces, such as a harmonic potential or a time-dependent force. How to build an effective Langevin equation to describe these forces and to analyze some important statistical quantities will be investigated in our future work.

\section*{Acknowledgments}
This work was supported by the National Natural Science Foundation of China under grant no. 11671182, and the Fundamental Research Funds for the Central Universities under grants no. lzujbky-2018-ot03 and no. lzujbky-2019-it17.

\appendix
\begin{widetext}
\section{Derivations of Eqs. \eqref{LWeq} and \eqref{LWt}}\label{App1}
Let us first derive Eq. \eqref{LWeq}, the Langevin equation of velocity process $v(t)$ evolving in physical time $t$. Replacing $s$ by $s(t)$ in the second equation of Eq. \eqref{LW_force}, one arrives at
\begin{equation}
\begin{split}
\frac{d}{d s(t)}v(s(t))=-\gamma v(s(t))+F \eta(s(t))+\xi(s(t)).
\end{split}
\end{equation}
It is equivalent to the equation
\begin{equation}
\begin{split}
\frac{d}{dt}v(t)&=-\gamma v(s(t))\frac{d}{dt}s(t)+F \eta(s(t))\frac{d}{dt}s(t)+\xi(s(t))\frac{d}{dt}s(t)\\
                &=-\gamma v(t)\frac{d}{dt}s(t)+F+\xi(s(t))\frac{d}{dt}s(t),
\end{split}
\end{equation}
where we have used the identity $v(t):=v(s(t))$ and
\begin{equation}
  \eta(s(t))=\frac{dt(s(t))}{ds(t)}= \frac{dt}{ds(t)}
\end{equation}
in the last line.

As for Eq. \eqref{LWt}, replacing $s$ in Eq. \eqref{LWs} with $s(t)$ gives
\begin{equation}
\begin{split}
v(t):&=v(s(t))\\
     &=F\int_0^{s(t)} e^{-\gamma(s(t)-s')}\eta(s')ds'+\int_0^{s(t)} e^{-\gamma(s(t)-s')}\xi(s')ds'\\
     &=F\int_0^t e^{-\gamma(s(t)-s(t'))}\eta(s(t'))ds(t')+\int_0^t e^{-\gamma(s(t)-s(t'))}\xi(s(t'))ds(t')\\
     &=F\int_0^t e^{-\gamma(s(t)-s(t'))}dt'+\int_0^t e^{-\gamma(s(t)-s(t'))}\xi(s(t'))ds(t'),
\end{split}
\end{equation}
where we use the variable substitution $s'\rightarrow s(t')$ in the third line.


\section{Derivations of Eq. \eqref{W3}}\label{App2}
Let us consider the symbol in Eq. \eqref{SymbolW}:
\begin{equation}\label{App2W}
 W(t_1,t_2,t_3,t_4)= e^{-\gamma s(t_1)}e^{\gamma s(t_2)}e^{-\gamma s(t_3)}e^{\gamma s(t_4)}
\end{equation}
for $t_4<t_3<t_2<t_1$. As Sec. \ref{five} discusses, the dominating role in $\langle W(t_1,t_2,t_3,t_4)\rangle$ comes from two parts; one is $p_0(t_4,t_1)$ denoting no renewal happens during $[t_4,t_1]$ and another one denoting some renewals happen in the middle interval $[t_3,t_2]$. The latter one depends on the four-point PDF $h(\cdot)$ containing $\delta(s_1-s_2)\Theta(s_2-s_3)\delta(s_3-s_4)$ of inverse subordinator $s(t)$.
Similarly to the two-point case, the expression $\langle W(t_1,t_2,t_3,t_4)\rangle$ can be calculated through Laplace transform ($t_1\rightarrow \lambda_1,t_2\rightarrow \lambda_2,t_3\rightarrow \lambda_3,t_4\rightarrow \lambda_4$)
\begin{equation}\label{Fourpoint}
\begin{split}
\langle W(\lambda_1,\lambda_2,\lambda_3,\lambda_4)\rangle
&=\int_0^\infty \int_0^\infty\int_0^\infty \int_0^\infty  e^{-\gamma s_1}e^{\gamma s_2}e^{-\gamma s_3}e^{\gamma s_4}h(s_1,\lambda_1;s_2,\lambda_2;s_3,\lambda_3;s_4,\lambda_4)ds_1ds_2ds_3ds_4.
\end{split}
\end{equation}
The key in Eq. \eqref{Fourpoint} is the four-point PDF $h(\cdot)$ of inverse subordinator $s(t)$, which is related to the corresponding four-point PDF $Z(\cdot)$ of subordinator $t(s)$ as
\begin{equation}
  h(s_1,\lambda_1;s_2,\lambda_2;s_3,\lambda_3;s_4,\lambda_4) = \frac{\partial}{\partial s_1}\frac{\partial}{\partial s_2}\frac{\partial}{\partial s_3}\frac{\partial}{\partial s_4}
    \frac{1}{\lambda_1\lambda_2\lambda_3\lambda_4} Z(\lambda_1,s_1;\lambda_2,s_2;\lambda_3,s_3;\lambda_4,s_4).
\end{equation}
Considering the independence and stationarity of the increments of subordinator $t(s)$, the four-point PDF $Z(\cdot)$ can be obtained as
\begin{equation}
  Z(\lambda_1,s_1;\lambda_2,s_2;\lambda_3,s_3;\lambda_4,s_4) = e^{-s_4\Phi(\lambda_1+\lambda_2+\lambda_3+\lambda_4)}e^{-(s_3-s_4)\Phi(\lambda_1+\lambda_2+\lambda_3)}
  e^{-(s_2-s_3)\Phi(\lambda_1+\lambda_2)}e^{-(s_1-s_2)\Phi(\lambda_1)}
\end{equation}
under the condition $s_4<s_3<s_2<s_1$. There are twenty-four different orders for $s_1,\cdots,s_4$, which makes the exact expression of $h(s_1,\lambda_1;s_2,\lambda_2;s_3,\lambda_3;s_4,\lambda_4)$ quite complicated. Fortunately, here we only need to know the term containing $\delta(s_1-s_2)\Theta(s_2-s_3)\delta(s_3-s_4)$ in $h(\cdot)$ which denotes some renewals happening in the middle interval $[t_3,t_2]$. This term can be obtained from totally four kinds of terms in $Z(\cdot)$, which are
\begin{equation}
\begin{split}
 &\Theta(s_1-s_2)\Theta(s_2-s_3)\Theta(s_3-s_4)e^{-s_4\Phi(\lambda_1+\lambda_2+\lambda_3+\lambda_4)}e^{-(s_3-s_4)\Phi(\lambda_1+\lambda_2+\lambda_3)}
  e^{-(s_2-s_3)\Phi(\lambda_1+\lambda_2)}e^{-(s_1-s_2)\Phi(\lambda_1)} \\
+&\Theta(s_1-s_2)\Theta(s_2-s_4)\Theta(s_4-s_3)e^{-s_3\Phi(\lambda_1+\lambda_2+\lambda_3+\lambda_4)}e^{-(s_4-s_3)\Phi(\lambda_1+\lambda_2+\lambda_4)}
  e^{-(s_2-s_4)\Phi(\lambda_1+\lambda_2)}e^{-(s_1-s_2)\Phi(\lambda_1)} \\
+&\Theta(s_2-s_1)\Theta(s_1-s_3)\Theta(s_3-s_4)e^{-s_4\Phi(\lambda_1+\lambda_2+\lambda_3+\lambda_4)}e^{-(s_3-s_4)\Phi(\lambda_1+\lambda_2+\lambda_3)}
  e^{-(s_1-s_3)\Phi(\lambda_1+\lambda_2)}e^{-(s_2-s_1)\Phi(\lambda_2)} \\
+&\Theta(s_2-s_1)\Theta(s_1-s_4)\Theta(s_4-s_3)e^{-s_3\Phi(\lambda_1+\lambda_2+\lambda_3+\lambda_4)}e^{-(s_4-s_3)\Phi(\lambda_1+\lambda_2+\lambda_4)}
  e^{-(s_1-s_4)\Phi(\lambda_1+\lambda_2)}e^{-(s_2-s_1)\Phi(\lambda_2)}. \\
\end{split}
\end{equation}
Taking derivatives with respect to $s_4,s_3,s_2,s_1$ on these four terms, together with some technical calculations, we find the factor multiplied by $\delta(s_1-s_2)\Theta(s_2-s_3)\delta(s_3-s_4)$ in $h(\cdot)$ is
\begin{equation}\label{delta2}
  \frac{[\Phi(\lambda_1)+\Phi(\lambda_2)-\Phi(\lambda_1+\lambda_2)]\cdot [\Phi(\lambda_1+\lambda_2+\lambda_3)+\Phi(\lambda_1+\lambda_2+\lambda_4)-\Phi(\lambda_1+\lambda_2)-\Phi(\lambda_1+\lambda_2+\lambda_3+\lambda_4)]}
  {\lambda_1\lambda_2\lambda_3\lambda_4\, e^{(s_2-s_3)\Phi(\lambda_1+\lambda_2)}e^{s_3\Phi(\lambda_1+\lambda_2+\lambda_3+\lambda_4)}}.
\end{equation}
Substituting it into \eqref{Fourpoint} gives
\begin{equation}\label{APPW}
  \begin{split}
 \frac{[\Phi(\lambda_1)+\Phi(\lambda_2)-\Phi(\lambda_1+\lambda_2)]\cdot [\Phi(\lambda_1+\lambda_2+\lambda_3)+\Phi(\lambda_1+\lambda_2+\lambda_4)-\Phi(\lambda_1+\lambda_2)-\Phi(\lambda_1+\lambda_2+\lambda_3+\lambda_4)]}
  {\lambda_1\lambda_2\lambda_3\lambda_4\, \Phi(\lambda_1+\lambda_2)\Phi(\lambda_1+\lambda_2+\lambda_3+\lambda_4)}.
  \end{split}
\end{equation}
Performing inverse Laplace transform of \eqref{APPW}, we can obtain the term of $\langle W(t_1,t_2,t_3,t_4) \rangle$ which represents some renewals happening in the middle interval $[t_3,t_2]$.
For $1<\alpha<2$, it can be seen that the term \eqref{APPW} is far less than another term $p_0(t_4,t_1)$ in $\langle W(t_1,t_2,t_3,t_4) \rangle$ for small $\lambda_i, i=1,\cdots,4$, while for $0<\alpha<1$ the two terms are of the same order and the sum is
\begin{equation}
  \langle W(t_1,t_2,t_3,t_4) \rangle \simeq p_0(t_4,t_3)p_0(t_2,t_1),
\end{equation}
which is the result in Eq. \eqref{W3} after changing the symbol of time ($t_4\longrightarrow t_1',t_3\longrightarrow t_1,t_2\longrightarrow t_2',t_1\longrightarrow t_2$).



\section{Derivations of EAMSD in Eq. \eqref{EA} and TAMSD in Eq. \eqref{TA}}\label{App3}
According to the generalized Green-Kubo formula \cite{DechantLutzKesslerBarkai:2014,MeyerBarkaiKantz:2017}, one should first rewrite the velocity correlation function in Eqs. \eqref{VV2} and \eqref{VV} into the scaling form with two separate parts to obtain the EAMSD and TAMSD, i.e.,
\begin{equation}
\begin{split}
\langle v(t)v(t+\tau)\rangle&=\langle v(t)v(t+\tau)\rangle_1+\langle v(t)v(t+\tau)\rangle_2\\
&\simeq C_1t^{\nu_1-2}\phi_1\left(\frac{\tau}{t}\right)+C_2t^{\nu_2-2}\phi_2\left(\frac{\tau}{t}\right);
\end{split}
\end{equation}
and we assume that $\phi_1(q)\rightarrow c_1q^{-\delta_1}$ and $\phi_2(q)\rightarrow c_2q^{-\delta_2}$ as $q\rightarrow 0$.

For the case $0<\alpha<1$, the velocity correlation function can be written as
\begin{equation}
\begin{split}
\langle v(t)v(t+\tau)\rangle
=\frac{F^2}{\Gamma(1-\alpha)\Gamma(\alpha)}t^2\phi_1\left(\frac{\tau}{t}\right)+\frac{D}{\gamma}\frac{1}{\Gamma(1-\alpha)\Gamma(\alpha)}\phi_2\left(\frac{\tau}{t}\right),
\end{split}
\end{equation}
where
\begin{equation}
\begin{split}
\phi_1(q) &= (1+q)^2 B\left(\frac{1}{1+q};\alpha+2,1-\alpha\right) + \alpha B\left(\frac{1}{1+q};\alpha,1-\alpha\right) \\
 &~~~ - (1+\alpha)(1+q)B\left(\frac{1}{1+q};\alpha+1,1-\alpha\right) + \Gamma(\alpha)\Gamma(1-\alpha)(1-\alpha)^2(1+q)
\end{split}
\end{equation}
and
\begin{equation}
\begin{split}
\phi_2(q)=B\left(\frac{1}{1+q};\alpha,1-\alpha\right).
\end{split}
\end{equation}
When $q\rightarrow0$,  $\phi_1(q)$ converges to the constant $c_1=\Gamma(\alpha)\Gamma(3-\alpha)/2$ and $\phi_2(q)$ to $c_2=\Gamma(\alpha)\Gamma(1-\alpha)$. Then we can obtain the parameters needed: $C_1=\frac{F^2}{\Gamma(1-\alpha)\Gamma(\alpha)}$, $\nu_1=4$, $C_2=\frac{D}{\gamma}\frac{1}{\Gamma(1-\alpha)\Gamma(\alpha)}$, and $\nu_2=2$.

After getting the necessary parameters, according to the generalized Green-Kubo formula, the second moment of $x(t)$ can be obtained as
\begin{equation}
\begin{split}
\langle x^2(t)\rangle=\langle x^2(t)\rangle_1+\langle x^2(t)\rangle_2=2K_1t^{\nu_1}+2K_2t^{\nu_2}
\end{split}
\end{equation}
with $\langle x^2(t)\rangle_i=\int_0^{t} \int_0^{t}\langle v(t_1')v(t_2')\rangle_idt_1' dt_2'$ and $K_i=\frac{C_i}{\nu_i}\int_0^\infty (1+q)^{-\nu_i}\phi_i(q)dq$, $i=1,2$. Further  substituting the expressions of $C_i$, $\nu_i$ and $\phi_i$ into it, one has
$K_1=F^2(\alpha(1-\alpha)(2-\alpha)(3+\alpha)/144+(1-\alpha)^2/8)$ and $K_2=\frac{D(1-\alpha)}{2\gamma}$.
Then the second moment of the L\'{e}vy walk with constant force
\begin{equation}
\begin{split}
\langle x^2(t)\rangle=2K_1t^4+\frac{D(1-\alpha)}{\gamma}t^2
\end{split}
\end{equation}
and the EAMSD
\begin{equation}
\begin{split}
\langle x^2(t)\rangle-\langle x(t)\rangle^2
\simeq\frac{F^2\alpha(1-\alpha)(2-\alpha)(3+\alpha)}{72}t^4
\end{split}
\end{equation}
for large time $t$.

As for the TAMSD of the L\'{e}vy walk with constant force, we split it into three parts
\begin{equation}
\begin{split}
\langle \overline{\delta^2(\Delta)}\rangle&=\frac{1}{T-\Delta}\int_0^{T-\Delta} \langle (x(t+\Delta)-x(t))^2\rangle dt -\frac{1}{T-\Delta}\int_0^{T-\Delta} \langle x(t+\Delta)-x(t)\rangle^2dt\\
&=\langle \overline{\delta^2(\Delta)}\rangle_1+\langle \overline{\delta^2(\Delta)}\rangle_2-\langle \overline{\delta^2(\Delta)}\rangle_3,
\end{split}
\end{equation}
where
\begin{equation}
\begin{split}
\langle \overline{\delta^2(\Delta)}\rangle_{i}=\frac{1}{T-\Delta}\int_0^{T-\Delta} \langle (x(t+\Delta)-x(t))^2\rangle_{i}dt
\end{split}
\end{equation}
and
\begin{equation}\label{3}
\begin{split}
\langle \overline{\delta^2(\Delta)}\rangle_3=\frac{1}{T-\Delta}\int_0^{T-\Delta} \langle x(t+\Delta)-x(t)\rangle^2dt.
\end{split}
\end{equation}
By virtue of the generalized Green-Kubo formula, the first two terms can be obtained as
\begin{equation}\label{12}
\begin{split}
\langle \overline{\delta^2(\Delta)}\rangle_{i}
=\frac{2c_{i}C_{i}}{(\beta_{i}+1)(\nu_{i}-\beta_{i}-1)(\nu_{i}-\beta_{i})}T^{\beta_{i}}\Delta^{\nu_{i}-\beta_{i}},
\end{split}
\end{equation}
where $\beta_{i}$ is the exponent of second moment of velocity $\langle v^2(t)\rangle_{i}\propto t^{\beta_{i}}$ with $\beta_1=2$ and $\beta_2=0$.
After substituting the parameters into Eq. \eqref{12}, one has
\begin{equation}
\begin{split}
\langle \overline{\delta^2(\Delta)}\rangle_1=\frac{F^2(1-\alpha)(2-\alpha)}{6}T^2\Delta^2,~~~~~\langle \overline{\delta^2(\Delta)}\rangle_2=\frac{D}{\gamma}\Delta^2.
\end{split}
\end{equation}
Besides,  substituting the first moment of position $\langle x(t)\rangle_F= F\frac{(1-\alpha)}{2}t^2$ into Eq. \eqref{3} leads to
\begin{equation}
\begin{split}
\langle \overline{\delta^2(\Delta)}\rangle_3=\frac{F^2(1-\alpha)^2}{3}T^2\Delta^2.
\end{split}
\end{equation}
Finally, the TAMSD for large $T$ and $\Delta\ll T$ is
$\langle \overline{\delta^2(\Delta)}\rangle\simeq\frac{F^2(1-\alpha)\alpha}{6}T^2\Delta^2$.

For the case $1<\alpha<2$, the velocity correlation function is
\begin{equation}
\begin{split}
\langle v(t)v(t+\tau)\rangle&=\langle v(t)v(t+\tau)\rangle_1+\langle v(t)v(t+\tau)\rangle_2\\
&=\frac{F^2(\alpha-1)}{2-\alpha}t^{3-\alpha}\phi_1\left(\frac{\tau}{t}\right)+\frac{D}{\gamma}t^{1-\alpha}\phi_2\left(\frac{\tau}{t}\right),
\end{split}
\end{equation}
where
\begin{equation}
\begin{split}
\phi_1(q)= (1+q)^{2-\alpha}-\frac{1}{3-\alpha}(1+q)^{3-\alpha}+\frac{1}{3-\alpha}q^{3-\alpha}
\end{split}
\end{equation}
and
\begin{equation}
\begin{split}
\phi_2(q)=q^{1-\alpha}-(1+q)^{1-\alpha}.
\end{split}
\end{equation}
When $q\rightarrow0$, we have $\phi_1(q)\rightarrow\frac{2-\alpha}{3-\alpha}$ and $\phi_2(q)\rightarrow q^{1-\alpha}$. Then we obtain the parameters needed: $C_1=\frac{F^2(\alpha-1)}{2-\alpha}$, $\nu_1=5-\alpha$, $c_1=\frac{2-\alpha}{3-\alpha}$, $\beta_1=3-\alpha$, $K_1=\frac{F^2(\alpha-1)}{2(4-\alpha)(5-\alpha)}$, $C_2=\frac{D}{\gamma}$, $\nu_2=3-\alpha$, $c_2=1$, $\beta_2=0$, and $K_2=\frac{D(\alpha-1)}{\gamma(2-\alpha)(3-\alpha)}$.
Finally, the second moment and the EAMSD of $x(t)$ are
\begin{equation}
\begin{split}
\langle x^2(t)\rangle=\frac{F^2(\alpha-1)}{(4-\alpha)(5-\alpha)}t^{5-\alpha}+\frac{2D(\alpha-1)}{\gamma(2-\alpha)(3-\alpha)}t^{3-\alpha}
\end{split}
\end{equation}
and
\begin{equation}
\begin{split}
\langle x^2(t)\rangle-\langle x(t)\rangle^2
\simeq \frac{F^2(\alpha-1)}{(4-\alpha)(5-\alpha)}t^{5-\alpha}
\end{split}
\end{equation}
for large time $t$.
Similarly, the TAMSD for large $T$ and $\Delta\ll T$ is
\begin{equation}
\begin{split}
\langle \overline{\delta^2(\Delta)}\rangle&=\frac{F^2(\alpha-1)}{(3-\alpha)(4-\alpha)}T^{3-\alpha}\Delta^2+\frac{2D}{\gamma (2-\alpha)(3-\alpha)}\Delta^{3-\alpha}-\frac{F^2(\alpha-1)^2}{(2-\alpha)^2(5-2\alpha)}T^{4-2\alpha}\Delta^2\\
&\simeq \frac{F^2(\alpha-1)}{(3-\alpha)(4-\alpha)}T^{3-\alpha}\Delta^2.
\end{split}
\end{equation}

\end{widetext}

\section*{References}
\bibliographystyle{apsrev4-1}
\bibliography{ReferenceW}

\end{document}